\documentclass[11pt,a4paper]{article}
\usepackage{jcappub}
\usepackage{amsbsy,esint,mathrsfs,commath,amsfonts,mathtools,amsthm}
\usepackage[dvipsnames]{xcolor}
\usepackage{subfig}
\usepackage[english]{babel}
\usepackage{tensor}

\newcommand*\diff{\mathop{}\!\mathrm{d}}

\newcommand{\units}[1]{\; \text{#1}} % unidades sin cursiva y espaciadas

\title{Sensitivity forecasts for gravitational-wave detectors to dark matter decaying into gravitons}
    
 \date{\today}
 
\author[a]{Jose A. R. Cembranos}
\author[a]{and \'Alvaro Cendal}
\affiliation[a]{Departamento de F\'isica Te\'orica and IPARCOS, Facultad de Ciencias F\'isicas, Universidad Complutense de Madrid, \\ 
Plaza de las Ciencias 1, 28040 Madrid, Spain}
\emailAdd{acendal@ucm.es}
\emailAdd{cembra@fis.ucm.es}

\abstract{
Dark matter may not be perfectly stable, and its decay could generate distinctive gravitational-wave signatures. In this work, we present model-independent predictions for the stochastic gravitational-wave background arising from the decay of ultralight dark matter into gravitons. Within this framework, we forecast the sensitivity reach of current and forthcoming gravitational-wave detectors to such signals.
}

\subheader{\normalfont IPARCOS-UCM-25-053}

\begin{document}
\maketitle
\flushbottom

\tableofcontents

\section{Introduction}
The fundamental nature of dark matter (DM) remains an open question in physics. Despite overwhelming cosmological and astrophysical evidence for its existence, its microscopic properties are still widely unknown. A wide variety of candidates have been proposed. Weakly Interacting Massive Particles (WIMPs) remain a paradigmatic class of DM candidates because they arise naturally in many motivated extensions of the Standard Model and freeze out with the correct relic abundance for weak-scale interactions. Classic realizations include the neutralino in R-parity conserving supersymmetry~\cite{Goldberg1983,Ellis1984}, lightest Kaluza-Klein states in models with universal extra dimensions~\cite{Servant2003,Cheng2002}, and new weakly interacting modes emerging from brane-world constructions and related compactification scenarios~\cite{Cembranos2003a,Cembranos:2003fu,Alcaraz2003}. These frameworks predict testable signals in direct detection, indirect searches and at colliders, and provide a concrete connection between electroweak-scale physics and cosmological DM.

An alternative and increasingly studied paradigm is coherent bosonic DM, where the dark sector is described by a classical, oscillating field. This category includes light scalars such as the QCD axion and axion-like particles~\cite{Peccei1977,Weinberg1978,Wilczek1978,Arvanitaki2010,Cicoli2012,Marsh2016,Hu2000,Amendola2006,Suarez2014,Hui2017,Irastorza2018}, coherent vector fields (dark photons / hidden U(1) condensates)~\cite{Nelson2011,Arias2012,Graham2016,Agrawal2021,Cembranos2012,Cembranos2013,Cembranos:2016ugq, Chase:2023puj, Chase:2024wsq,Chase:2025wwj} and even massive spin-2 candidates (hidden graviton)~\cite{Cembranos:2013cba,Aoki:2017cnz, Marzola:2017lbt,Cembranos2017,Cembranos2022}. These coherent bosonic modes can reproduce cold-dark-matter behaviour at large scales while producing distinctive small-scale structure (e.g. solitonic cores, cutoffs in the power spectrum) and novel observational signatures. The work we develop here focuses on these coherent bosonic frameworks and tests its possible decay into gravitons. 

In fact, given the absence of a clear theory, the study of DM phenomenology often relies on model-independent approaches that explore possible interactions, production mechanisms and decay channels without committing to a specific particle framework. In this context, the possibility of a unstable DM particle has been considered in the literature \cite{Cembranos:2005us,CembranosEtAl2007,DeLopeAmigoEtAl2009, IbarraEtAl2013, PoulinEtAl2016}. Decaying DM faces several observational constraints, mainly from large scale structure (LSS) formation \cite{PoulinEtAl2016, WangZentner2012}. However, it has been shown that different proposed models of decaying DM may mitigate some astrophysical and cosmological problems \cite{Cembranos:2005us,CembranosEtAl2007,WangEtAl2014,PandeyEtAl2020}. 

Although the main focus has been on decay channels producing Standard Model particles (mainly photons, positrons and neutrinos), the possibility of DM decaying into gravitons has also been considered in the literature \cite{Alonzo-ArtilesEtAl2021,EmaEtAl2022a,LandiniStrumia2025}, showing that couplings to quadratic invariants of the Riemann tensor lead to gravitational decays. In this work, we will study the phenomenology an unstable ultralight DM particle $\phi$ with mass $m_\phi$ whose dominant decay channel is into one or more gravitons. For simplicity, we will assume that the decay is into two massless gravitons, $\phi \rightarrow 2h$, although more general decays give qualitative similar results. In a very different context, the phenomenology of these kinds of decays has been recently studied in \cite{Ramazanov:2023nxz, DunskyEtAl2025a} using the inverse Gertsenshtein mechanism ---the conversion of gravitons into photons in the presence of magnetic fields--- to give constraints on the DM lifetime from X-ray and gamma-ray observations. Here, we instead focus on the direct gravitational wave (GW) signal produced by the decay, giving a forecast for the detectability of this process in current and future GW detectors. 

Since the detection of the first GW signal by LIGO in 2015 \cite{LV_GW150914}, the field of GW astronomy has rapidly expanded. Ground-based interferometers such as LIGO, Virgo and KAGRA (LVK) are sensitive to the Hz--kHz range, where they have already detected numerous compact binary mergers \cite{AbbottEtAl2023}. At much lower frequencies, pulsar timing arrays (PTAs) like NANOGrav and the International Pulsar Timing Array (IPTA) have reported evidence for a stochastic gravitational wave (GW) background (SGWB) in the nHz band \cite{NANOGrav15}, a window that will be greatly improved with the Square Kilometre Array (SKA) \cite{JanssenEtAl2014}. In the coming decades, space-based interferometers will explore the intermediate mHz--deci-Hz regime: LISA \cite{LISA2024}, already approved by ESA, will target the mHz range, while concepts such as DECIGO \cite{DECIGO2021} and BBO \cite{BBO2009} are designed for the deci-Hz band. Together with third-generation ground-based detectors like the Einstein Telescope (ET) \cite{ET2025} and Cosmic Explorer (CE) \cite{CE2021}, this network of experiments will cover frequencies from nHz to kHz, providing a unique opportunity to test exotic sources of GWs.

The manuscript is organized as follows. Section \ref{sec:SGWB_detection} provides a brief overview of SGWB detection strategies, focusing on interferometers and pulsar timing arrays (PTAs). Section \ref{sec:DM_decay} introduces the theoretical framework for the production of a SGWB from DM decay. Section \ref{sec:results} will present our main results, including forecasts for the detectability of the signal for current (LVK and IPTA) and future (LISA, ET, CE, BBO and SKA) detectors. Finally, Section \ref{sec:conclusions} will summarize our conclusions. Throughout this work\footnote{\textit{Notation and conventions:} Lowercase Latin indices $i,j,k,...$ run over spatial coordinates $1,2,3$. $A,B,...$ denote the two polarization states of a GW, $+,\times$. $I, J,...$ label different detectors. Apart from using energy units to express masses (i.e. each mass must be understood as $m c^2$), we will not be using natural units.
}, we will assume a flat $\Lambda$CDM cosmology with parameters taken from the latest Planck results \cite{Planck2018}.

\section{Detection of stochastic GW backgrounds}\label{sec:SGWB_detection}
A SGWB, i.e., a random, persistent GW signal resulting from the superposition of numerous unresolved and independent sources, requires different detection techniques than those used for transient GW signals. The most common approach to detect a SGWB is through cross-correlation of the outputs of multiple detectors, allowing to distinguish the common GW signal from uncorrelated noise. In this section, we will introduce the definition of signal-to-noise ratio (SNR) for SGWB detection via cross-correlation \cite{AllenRomano1999}, which we will use in Sec. IV to forecast the detectability of the signal from DM decay.

\subsection{Stochastic GWs}
In the transverse-traceless gauge, the metric perturbation $h_{ij}(t, \vec{x})$ associated to a GW can be expressed as 
\begin{align}
    h_{ij}(t,\vec{x}) = \! \int_{-\infty}^{\infty} \hspace{-5pt} \diff f \int_{S^2} \! \diff \Omega_{\hat{u}} \hspace{-5pt} \sum_{A=+,\times} \hspace{-5pt} \tilde{h}_A(f,\hat{u}) e^A_{ij}(\hat{u}) e^{2\pi i f(t - \hat{u} \cdot \vec{x}/c)},
\end{align}
where $e^A_{ij}(\hat{u})$ are the polarization tensors for the two independent polarization states $A = +, \times$, $\hat{u}$ is the normal vector pointing in the direction of propagation of the wave, and $\tilde{h}_A(f,\hat{u})$ are the Fourier components of the metric perturbation with frequency $f$. 

For a SGWB, we assume that the Fourier components are random variables. We may assume without loss of generality that they have zero mean, $\langle h_A(f,\hat{u}) \rangle = 0$. For an isotropic, unpolarized and stationary background, the quadratic expectation values of the Fourier components can be expressed as \cite{ThraneRomano2013}
\begin{equation}
    \label{eq:SGWB_correlation}
    \langle \tilde{h}_A(f,\hat{u}) \tilde{h}_{A'}^*(f',\hat{u}') \rangle = \frac{1}{16 \pi} \delta(f - f') \delta_{AA'} \delta^2(\hat{u},\hat{u}') S_h(f),
\end{equation}
where $S_h(f)$ is the one-sided power spectral density (PSD) of the SGWB. The PSD is related to the dimensionless energy density parameter $\Omega_{\text{GW}}(f)$, the characteristic strain $h_c(f)$ and the spectral energy density $S_E(f)$ as follows:
\begin{equation}
    \label{eq:relations}
    H_0^2 \Omega_{\text{GW}}(f) = \frac{2\pi^2}{3} f^3 S_h(f) = \frac{2\pi^2}{3} f^2 h_c^2(f) = \frac{8 \pi G}{3 c^2} f S_E(f),
\end{equation}
where $H_0 = 67.66 \units{km/s/Mpc}$ is the Hubble parameter today. While for transient signals the characteristic strain $h_c(f)$ is the most common quantity used in the literature, for SGWB the energy density parameter $\Omega_{\text{GW}}(f)$ is more frequently employed. 

\subsection{Detector response and cross-correlation}
The response $h(t)$ of a GW detector to an incoming GW is a convolution of the metric perturbation $h_{ij}$ with the detector-response function $\mathcal{R}_{ij}$. In the frequency domain, this can be expressed as
\begin{equation}
    \tilde{h}(f) = \int_{S^2} \! \diff \Omega_{\hat{u}} \hspace{-5pt} \sum_{A=+,\times} \hspace{-5pt} \tilde{h}_A(f,\hat{u}) \mathcal{R}^A(\hat{u},f) e^{-i 2 \pi f \hat{u} \cdot \vec{x}/c},
\end{equation}
where $\mathcal{R}^A(\hat{u},f)$ is the frequency-domain response function for polarization $A$ with frequency $f$ and direction $\hat{u}$. 

To search for SGWBs, we are interested in the correlation between the outputs of different detectors $I,J$, defined as 
\begin{equation}
    \langle \tilde{h}_I(f) \tilde{h}_J^*(f') \rangle = \frac{1}{2} \delta(f - f') \Gamma_{I \! J}(f) S_h(f),
\end{equation}
where the overlap reduction function (ORF) $\Gamma_{I \! J}(f)$ encodes the relative geometry and separation of the detectors. Explicitly,
\begin{equation}
    \label{eq:ORF}
    \Gamma_{I \! J}(f) = \frac{1}{8\pi} \int_{S^2} \! \diff \Omega_{\hat{u}} \hspace{-5pt} \sum_{A=+,\times} \hspace{-5pt} \mathcal{R}_I^A(\hat{u},f) \mathcal{R}_J^{A*}(\hat{u},f) e^{-i 2 \pi f \hat{u} \cdot \Delta \vec{x}/c},
\end{equation}
with $\Delta \vec{x} = \vec{x}_I - \vec{x}_J$ the separation vector between the two detectors. 
It is common to introduce the normalized ORF $\gamma_{I \! J}(f)$, defined such that $\gamma_{I \! J}(f) = 1$ for two identical and co-located detectors.  This normalization separates the effects of detector geometry and relative orientation from the individual sensitivity of each instrument. For two interferometers with opening angles $\alpha_I$ and $\alpha_J$, the relation between the normalized and unnormalized ORFs is
\begin{equation}
    \Gamma_{I \! J}(f) =\frac{\sin \alpha_I \sin \alpha_J}{5} \gamma_{I \! J}(f).
\end{equation}

We can also extend this treatment to PTAs, where the ORF between two pulsars separated by an angle $\theta_{I \! J}$ is given by~\cite{ThraneRomano2013}
\begin{equation}
    \Gamma_{I \! J} = \frac{1}{2 \pi f^2} \frac{1}{3} \zeta_{I \! J},
\end{equation}
where $\zeta_{I \! J}$ are the Hellings-Downs factors,
\begin{align}
    \zeta_{I \! J} = \frac{3}{2} \left( \frac{1 - \cos \theta_{I \! J}}{2} \right) \log \left( \frac{1 - \cos \theta_{I \! J}}{2} \right) - \frac{1}{4} \left( \frac{1 - \cos \theta_{I \! J}}{2} \right) + \frac{1}{2} + \frac{1}{2} \delta_{I \! J}. 
\end{align}
\subsection{Signal-to-noise ratio}
\begin{figure}
    \centering
    \includegraphics[width=0.75\linewidth]{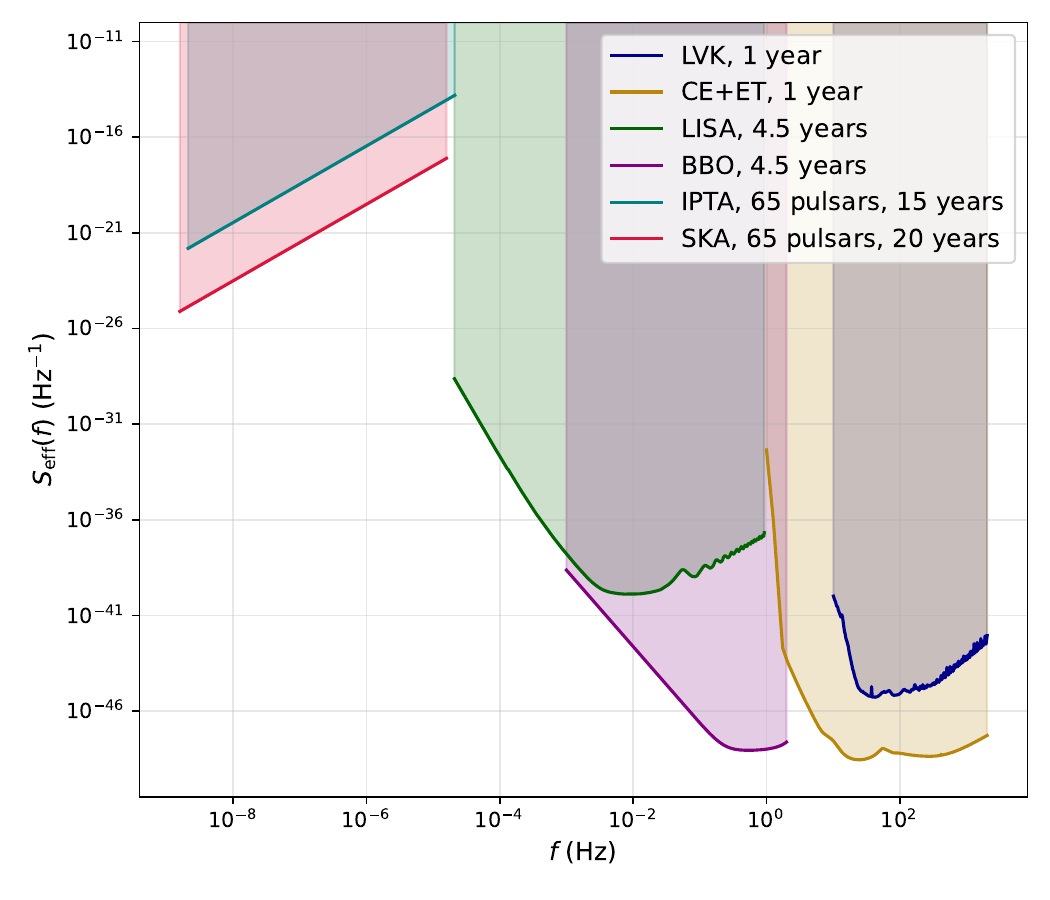}
    \caption{Effective noise PSD $S_{\text{eff}}(f)$ as a function of frequency for the detectors considered in this work.}
    \label{fig:Seff}
\end{figure}

We will be using the signal-to-noise ratio (SNR) as a figure of merit for the detectability of a SGWB. The optimal SNR from the cross-correlation of $N$ detectors after an observation time $T$ is given by \cite{Maggiore2007,AllenRomano1999}
\begin{equation}
    \text{SNR}= \sqrt{2 T} \left[ \int_0^\infty \! \diff f  \sum_{I = 1}^N \sum_{J > I}^N \frac{\Gamma_{I \! J}^2(f) S_h^2(f)}{P_{nI}(f) P_{nJ}(f)} \right]^{1/2},
\end{equation}
where $P_{nI}(f)$ is the noise PSD of detector $I$. We will consider as a detection criterion $\text{SNR} \geq 8$ in order to give a conservative estimate of the detectability of the signal. We may define an effective noise PSD $S_{\text{eff}}(f)$ as
\begin{equation}
    \label{eq:Seff}
    S_{\text{eff}}(f) = \left[\sum_{I = 1}^N \sum_{J > I}^N \frac{\Gamma_{I \! J}^2(f)}{P_{nI}(f) P_{nJ}(f)}\right]^{-1/2}.
\end{equation}
This way, we can rewrite the SNR as
\begin{equation}
    \label{eq:SNR}
    \text{SNR} = \sqrt{2 T} \left[ \int_0^\infty \! \diff f  \frac{S_h^2(f)}{S_{\text{eff}}^2(f)} \right]^{1/2}.
\end{equation}
In this work we will consider six cases: \textit{i)} the current LIGO-Virgo-KAGRA (LVK) network, \textit{ii)} the future ET+CE network formed by Einstein Telescope (ET) and Cosmic Explorer (CE), \textit{iii)} LISA, \textit{iv)} BBO, \textit{v)} the current IPTA network and \textit{vi)} the future SKA observatory. In Fig. \ref{fig:Seff} we show the effective noise PSD $S_{\text{eff}}(f)$ for each of these cases.

\begin{figure}
    \centering
    \includegraphics[width=0.75\linewidth]{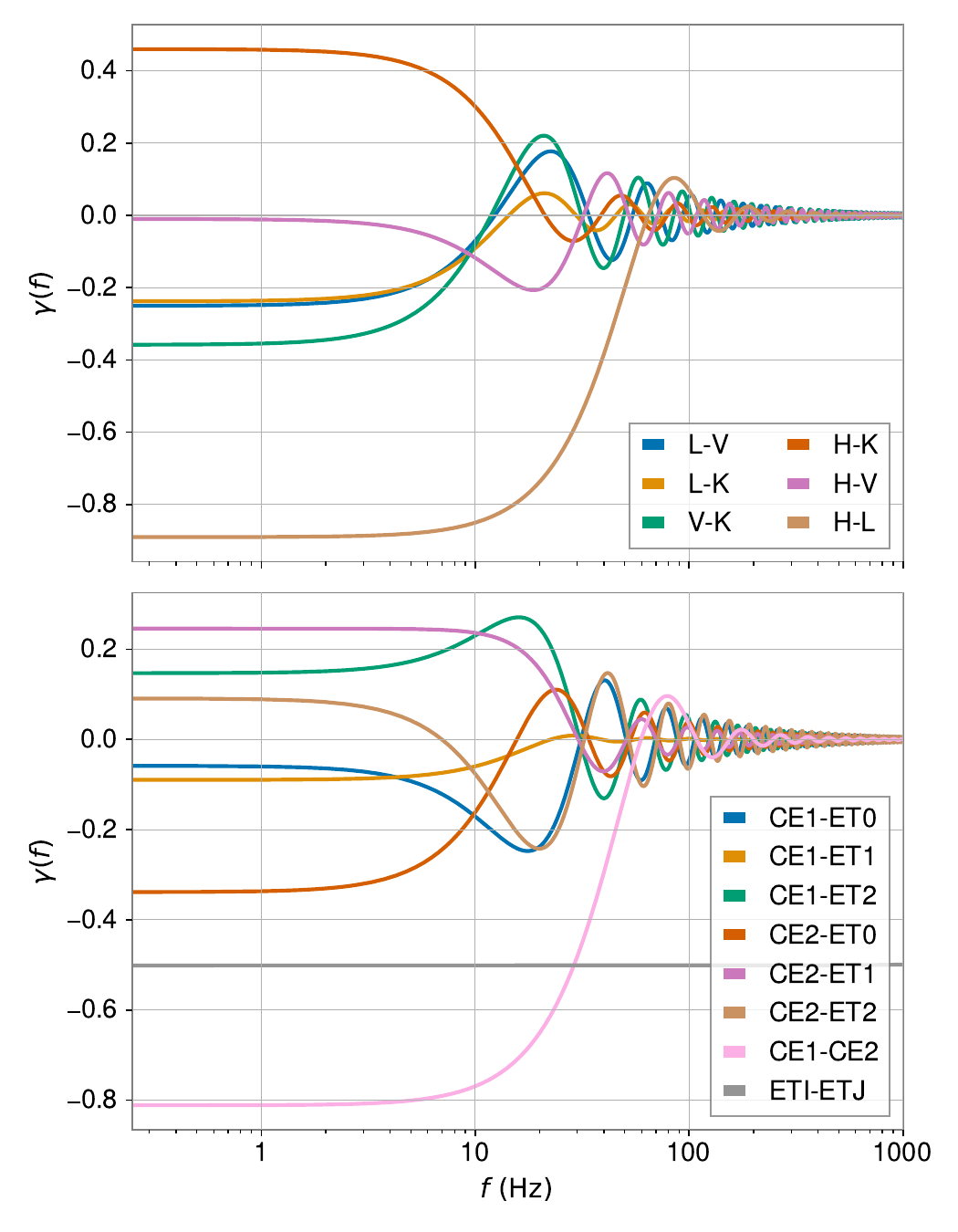}
    \caption{Normalized overlap reduction functions $\gamma_{I \! J}(f)$ as a function of frequency for the different pairs of detectors in the LVK and ET-CE networks.}
    \label{fig:ORFs}
\end{figure}

For ground-based interferometers, we consider the LVK network formed by the two LIGO detectors (Hanford and Livingston), Virgo and KAGRA, evaluating the optimal SNR for the full network. We will also consider the prospective third-generation detectors, ET and CE, treated as a separate network. To compute the ORFs for LVK and CE-ET, shown in Fig. \ref{fig:ORFs}, we used the \texttt{pygwb} package \cite{pygwb}. For the ET--CE overlap reduction function, we adopt the fiducial locations and orientations specified in \cite{GuptaEtAl2024}, with two CE observatories (CE-A, with 40km length, and CE-B, with 20km length). Combining LVK with ET-CE does not lead to a significant improvement in the overall SNR.
For LISA, one could in principle attempt to cross-correlate the noise-orthogonal  time-delay interferometry (TDI) channels $A$, $E$, and $T$. Since the $T$ is largely insensitive to GWs, it is well suited for monitoring instrumental noise. However, for frequencies $f \ll 3 \times 10^{-2} \units{Hz}$, the $A$ and $E$ channels have uncorrelated responses to an unpolarized, isotropic SGWB \cite{RomanoCornish2017}. We will result to the autocorrelation on a single channel. Assuming that the instrumental noise can be perfectly subtracted (with $T$ acting as a noise monitor), the corresponding ideal SNR can be written as \cite{ThraneRomano2013}
\begin{equation}
    \label{eq:SNR_LISA}
    \text{SNR} = \sqrt{T} \left[ \int_0^\infty \! \diff f  \frac{S_h^2(f)}{S_n^2(f)} \right]^{1/2},
\end{equation}
where $S_n = P_n / \mathcal{R}$ and $\mathcal{R}$ is the transfer function of the detector. The missing factor of $\sqrt{2}$ with respect to Eq. \eqref{eq:SNR} comes from the fact that we are only using one channel.

For BBO, which consists of two space-based interferometers, we use the ORF and noise PSD provided in the supplementary material of \cite{ThraneRomano2013}.

For PTAs, we consider two cases: (i) the current IPTA network, modeled with a RMS timing noise of $100 \units{ns}$, a cadence of $\Delta t = 2 \units{weeks}$, 65 pulsars from the DR2 catalog \cite{IPTADR2}, and a total observation time of 15 years; and (ii) the future SKA observatory, using the same pulsar set and cadence but with an improved RMS timing noise of $20 \units{ns}$ and a total duration of 20 years. We have checked that the results remain essentially unchanged when varying the pulsar distribution or reducing the total number of pulsars.

\section{SGWB from DM decay}\label{sec:DM_decay}
The specific signal we will be considering in this work, in order to give numerical concrete results, is a stochastic gravitational-wave background (SGWB) produced by the decay of bosonic (not necessarily scalar) DM particles $\phi$ into two gravitons, $\phi \rightarrow 2h$. This process has been discussed in the literature \cite{Alonzo-ArtilesEtAl2021, EmaEtAl2022a, LandiniStrumia2025, Ramazanov:2023nxz, DunskyEtAl2025a}, and can arise from the linearization of exotic couplings to the Kretschmann and Chern-Pontryagin invariants, such as those of Chern-Simons modifications of General Relativity \cite{JackiwPi2003} or scalar Gauss-Bonnet theories \cite{FernandesEtAl2022}. 
Decays into gravitons are also allowed at loop level within the Standard Model, though they are extremely suppressed by the Planck scale \cite{DelbourgoLiu2001}. Extensions of the Standard Model, including scenarios with extra dimensions, can introduce couplings to gravity that are not Planck-suppressed, potentially leading to shorter lifetimes \cite{ObiedEtAl2024}.

Rather than focusing on a specific particle physics model, we adopt a fully phenomenological approach, characterizing the process by the DM mass $m_\phi$ and lifetime $\tau_\phi$, which are enough to fully parametrize the signal. We further assume that the gravitational decay channel is dominant over other possible decay channels, and that this decaying DM particle constitutes a fraction $r_{\text{DDM}}$ of the total DM today, the rest being stable particles. This approach allows us to provide sensitivity forecasts applicable to any models fulfilling these criteria, highlighting the potential observational consequences of graviton decays without being tied to any particular model.

Previous work on decaying DM has heavily constrained the lifetime of the decaying particle, as any noticeable change in the DM abundance affects the LSS formation. More precisely, in \cite{PoulinEtAl2016} the authors study model-independent decays to dark radiation, constraining the lifetime of the decaying DM to be $\tau_\phi \cdot r_{\text{DDM}} > 170 \units{Gyr} = 12.3 \; t_0$, where $t_0 = 13.8 \units{Gyr}$ is the present age of the Universe. Given this constraint, we will be considering lifetimes well above the age of the universe, such that the modification to the standard $\Lambda$CDM cosmology is negligible and structure formation does not play any role in the analysis. Regarding the mass of the DM particle, recent work has constrained the mass of bosonic DM candidates to be $m_\phi > 2.2 \times 10^{-21} \units{eV}$ \cite{ZimmermannEtAl2025}, which we will also be taking into account.

The contribution to the SGWB from DM decay can be divided into two main components: extragalactic and galactic. The extragalactic component originates from the decay of the cosmological DM density and can be treated as isotropic. The galactic component comes from the decay of the Milky Way DM halo, which is anisotropic and depends on the position of the observer within the halo. In the following subsections, we will describe each contribution in detail.

\subsection{Extragalactic contribution}
The flux of gravitons today from a $\phi \rightarrow 2h$ decay due to a homogeneous and isotropic distribution of DM is given by \cite{CembranosEtAl2007}
\begin{align}
    \label{eq:extragalactic_flux}
    \frac{\diff \Phi_{\text{E}}}{\diff E}  &= 2 c \int^{t_0}_0 \frac{n_\phi(t)}{\tau_\phi}  \delta(E - a(t) m_\phi/2) \diff t \nonumber \\
                                &= 2c \frac{n_{\phi}(t_0)}{\tau_\phi} \frac{e^{-(t(a = 2E/m_\phi)-t_0)/\tau}}{E \cdot H(2E/m_\phi)} \Theta \left(\frac{m_\phi}{2} - E \right),
\end{align} 
where $n_\phi(t)$ is the (comoving) number density of $\phi$ particles at time $t$, $a(t)$ is the scale factor, $t(a)$ is the time corresponding to scale factor $a(t)$, $H(a)$ is the Hubble parameter and $E$ is the energy of the graviton today. The factor of $2$ accounts for the two gravitons produced in each decay.
We will assume that the decaying DM is not produced nor decays in any other channel, such that we can express the number density of decaying DM particles as
\begin{equation}
    n_\phi(t) = n_\phi(t_0) e^{-(t-t_0)/\tau_\phi} = \frac{r_{\text{DDM}}\rho_c \Omega_{\text{DM}}}{m_\phi} e^{-(t-t_0)/\tau_\phi},
\end{equation}
where $\rho_c$ is the critical density of the universe and $\Omega_{\text{DM}}$ is the total present DM density parameter.

To establish a connection between this flux $\Phi$, and the corresponding SGWB signal, we rewrite the graviton number density
\begin{equation}
    n_h = \Phi / c = \int \frac{1}{c} \frac{\diff \Phi}{\diff E} \diff E ,
\end{equation}
so the energy density is given by 
\begin{equation}
    \rho_h = \int \frac{2\pi \hbar}{c} \left(E \frac{\diff \Phi}{\diff E} \right) \diff f.
\end{equation}
By definition, the integrand is the spectral energy density $S_E(E)$, so using Eq. \eqref{eq:relations} we can write the SGWB quantities in terms of the flux of gravitons as
\begin{align}
    H_0^2 \Omega_{\text{GW}}(f)  &= \frac{16 \pi^2 G}{3 c^3} \hbar f \left(E \frac{\diff \Phi}{\diff E} \right), \\
    S_{h} &= \frac{8 \hbar G}{c^3} \frac{1}{f^2} \left(E \frac{\diff \Phi}{\diff E} \right), \label{eq:S_h_E}\\
    h_{c} &= \sqrt{ \frac{8 \hbar G}{c^3} \frac{1}{f} \left(E \frac{\diff \Phi}{\diff E} \right)}. \label{eq:h_c_E}
\end{align}
This way, we have established a connection between the flux of gravitons from DM decay and the corresponding SGWB quantities, which we will use to assess the detectability of the signal.
\subsection{Local contribution}
The local contribution to the SGWB has to be handled with more care. Even though most DM halo profiles are spherically symmetric, the position of the observer (i.e., our position in the Milky Way) breaks this symmetry, leading to an anisotropic signal. The flux of gravitons per unit solid angle from a direction characterized by an angle $\psi$ with respect to the Galactic Center is given by 
\begin{equation}
    \frac{\diff \Phi_{\text{L}}}{\diff \Omega} = \frac{1}{4 \pi} \int^{\infty}_{0} \frac{\diff n_h[r(s,\psi)]}{\diff t} \diff s,
\end{equation}
where
\begin{equation}
    \frac{\diff n_h (r)}{\diff t} = \frac{2 \, r_{\text{DDM}} \, \rho(r)}{\tau_\phi m_\phi}
\end{equation}
is the production rate of gravitons per unit volume at a distance $r$ from the Galactic Center, $\rho(r)$ is the DM mass density profile and $s$ is the distance along the line of sight. The effect of the decay on the DM density, which should be accounted for when computing the integral to all distances, is negligible for $\tau_{\phi} \gg t_0$ (for a halo radius $s \sim \mathcal{O} (100 \! \units{kpc})$, the factor due to decay is $\exp(s/c\tau) \sim \exp(10^{-5}) \simeq 1$).

Neglecting this factor, we can express the total flux as
\begin{equation}
    \Phi_{\text{L}} = \frac{r_{\text{DDM}}}{m_\phi \tau_\phi} \int^\infty_0 \rho(r) \frac{r}{R} \log \left(\frac{1 + r/R}{|1-r/R|}\right) \diff r,
\end{equation}
where $R = 8.2 \units{kpc}$ is the distance from the observer (i.e. the Solar System) to the Galactic Center. In this work we will be using the Einasto profile \cite{OuEtAl2024},
\begin{equation}
    \rho(r) = \frac{M_0}{4 \pi r_s^3} \exp\left[-(r/r_s)^\alpha\right],
\end{equation}
where $M_0$, $r_s$ and $\alpha$ are parameters fitted to the Milky Way halo. In \cite{OuEtAl2024} the authors find that the Einasto profile with $M_0 = 0.62 \times 10^{11} M_{\odot}$, $r_s = 3.86 \units{kpc}$ and $\alpha = 0.91$ gives a better fit to the Milky Way halo than the more commonly used Navarro-Frenk-White (NFW) profile. We will be using these values in the following, although we don't expect significant changes if other sensible profiles are used.

Given the proximity of the source, the redshift of the gravitons produced in the halo is negligible. Nevertheless, due to the velocity dispersion of the DM particles in the halo, the signal will not be perfectly monochromatic. For velocity dispersions $\sigma_v \sim 280 \units{km/s}$ \cite{StaudtEtAl2024}, the corresponding Doppler broadening is $\Delta E / E \simeq \sigma_v / c  \simeq 10^{-3}$. We will thus characterize the local contribution as a sharply peaked signal around $E_h = m_\phi / 2$ with a width $\sigma_E \simeq 10^{-3}\, E_h$,
\begin{equation}
    \label{eq:galactic_flux}
\frac{\diff \Phi_{\text{L}}}{\diff E} = \Phi_{\text{L}} \frac{1}{\sqrt{2 \pi} \sigma_E} \exp \left[-\frac{(E - m_\phi/2)^2}{2 \sigma_E^2}\right].
\end{equation}
When considering this kind of signal, we must take into consideration the frequency resolution of the detector. To do so, we will consider $\sigma_f = \sigma_E / (2 \pi \hbar) = \max (2 \cdot 10^{-3} f_h, 1/T_{\text{obs}})$, where  $f_h = E_h/(2 \pi \hbar)$ and $T_{\text{obs}}$ is the observation time.

Note that applying the method described in Sec. \ref{sec:SGWB_detection} to the differential flux \eqref{eq:galactic_flux} will not take into account the anisotropy of the local contribution. The full treatment of an anisotropic SGWB is beyond the scope of this work (see \cite{ThraneEtAl2009, CusinEtAl2017} and references therein for more details). In this study, we provide a first estimate of the galactic contribution by treating it as isotropic, which yields a conservative assessment of the signal detectability. Given its characteristic spatial distribution, a fully anisotropic treatment would be expected to enhance the results of this analysis. Owing to the directional sensitivity of PTAs, we will restrict our consideration of the galactic contribution to interferometric detectors.

\subsection{Combining extragalactic and local contributions}
\begin{figure}
    \centering
    \includegraphics[width=0.75\linewidth]{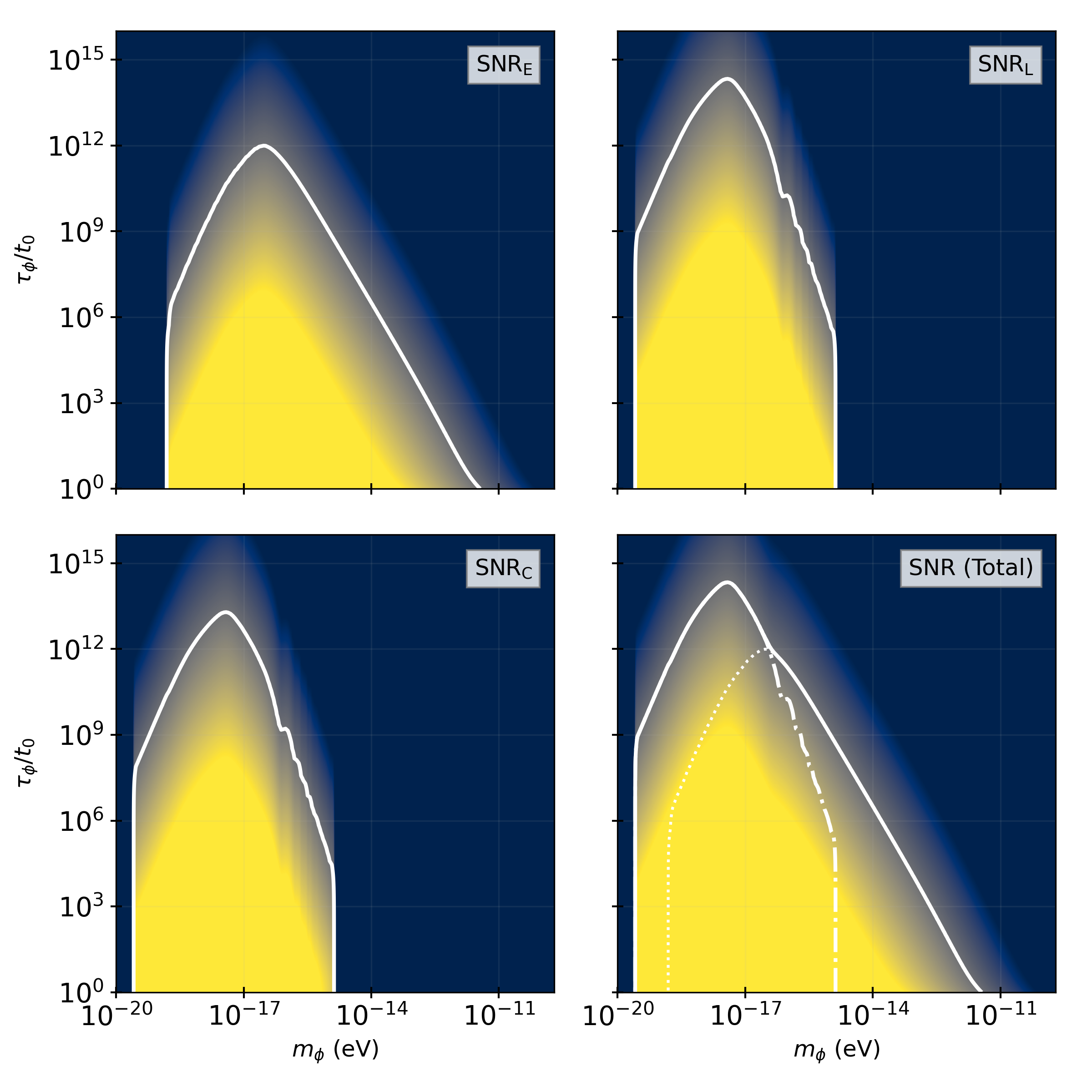}
    \caption{Breakdown of the different contributions to the total SNR for LISA (from left to right, top to bottom: extragalactic, local, cross-term and total SNR) in the $(m_\phi, \tau_\phi)$ parameter space. Solid lines show the detectability bound $\text{SNR} = 8$.  Dashed and dot-dashed lines show the extragalactic and local contributions to total, respectively. \label{fig:LISA_forecast}}
\end{figure}

The total SGWB signal from DM decay is the sum of the extragalactic and local contributions. Given that both contributions are independent, we can express the total flux as $\Phi = \Phi_{\text{L}} + \Phi_{\text{E}}$. This linearity carries over to the SGWB quantities, but not to the SNR, which depends quadratically on the PSD. Thus, we can express the total SNR as a sum of local (L) and extragalactic (E) contributions, plus a cross-term (C): 
\begin{equation}
    \text{SNR}^2 = \text{SNR}_{\text{L}}^2 + \text{SNR}_{\text{E}}^2 + \text{SNR}^2_{\text{C}},
\end{equation} 
where $\text{SNR}_{\text{E}}$ and $\text{SNR}_{\text{L}}$ are given by the usual expression \eqref{eq:SNR} (\eqref{eq:SNR_LISA} for LISA) with $S_h(f)$ \eqref{eq:S_h_E} evaluated at the extragalactic and local fluxes, respectively. The cross-term arises from the product of the local and extragalactic PSDs in the integrand of Eq. \eqref{eq:SNR},
\begin{equation}
    \label{eq:SNR_cross}
    \text{SNR}_{\text{C}}^2 = A \cdot T \left[ \int_0^\infty \! \diff f  \frac{S_{h,\text{L}}(f) S_{h,\text{E}}(f)}{S_{\text{eff}}^2(f)} \right],
\end{equation}
where $A = 4$ for correlation between two detectors and $A = 2$ for single-detector autocorrelation (as in LISA). As an example, in Fig. \ref{fig:LISA_forecast} we show the breakdown of the different contributions to the total SNR for LISA, considering $r_{\text{DDM}} = 1$. The local contribution dominates at low masses, while the extragalactic contribution takes over at high masses. The cross-term contribution is always subdominant.

\section{Results}\label{sec:results}
\begin{figure}[h]
    \centering
    \includegraphics[width=\linewidth]{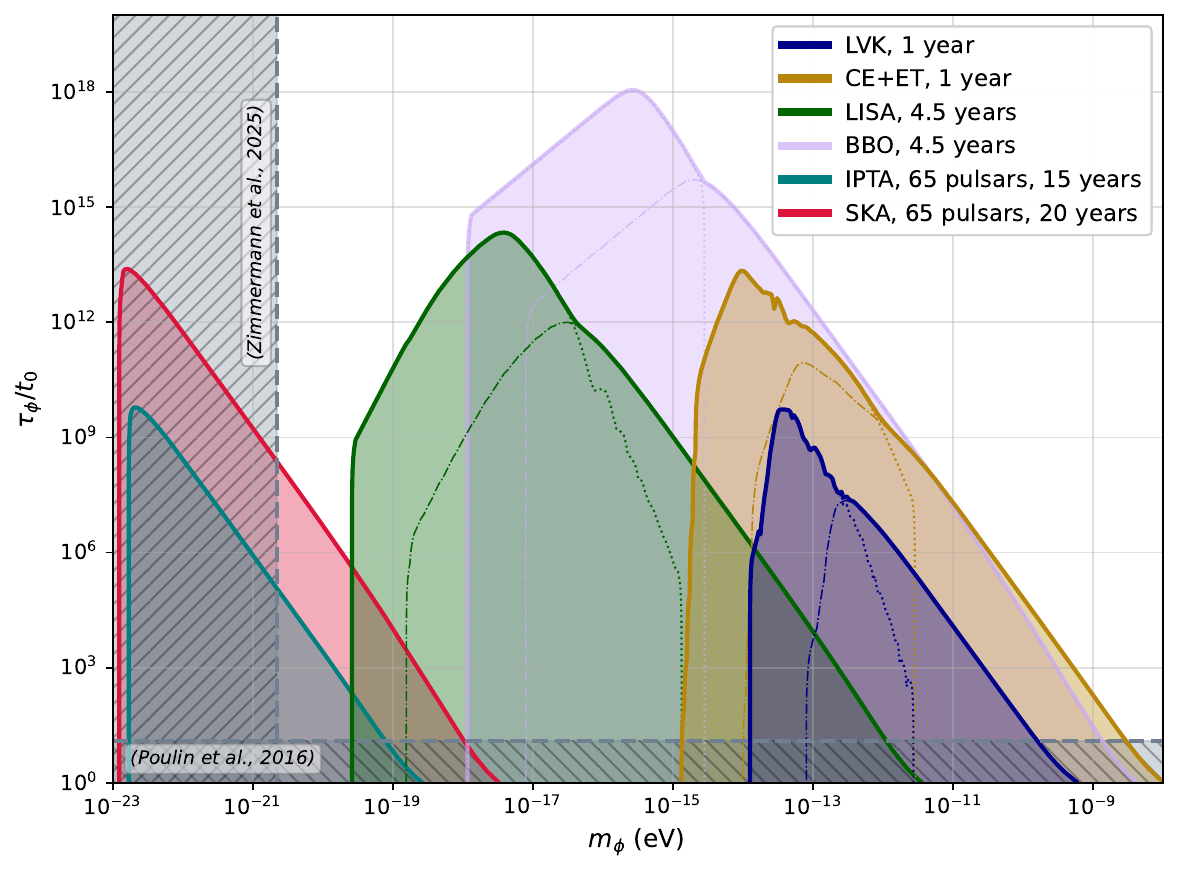}
    \caption{Forecast for the detectability ($\text{SNR} = 8$, solid line) of a SGWB from DM decay into two gravitons, $\phi \rightarrow 2h$, in the $(m_\phi, \tau_\phi)$ parameter space for different current and future GW detectors. Dotted and dash-dotted lines show the local and extragalactic contributions, respectively. Hatched regions show constraints from previous works.
    \label{fig:forecast}
    }
\end{figure}

Figure \ref{fig:forecast} shows the forecasted detectability of the SGWB from DM decay to gravitons, $\phi \to 2h$ in the $(m_\phi, \tau_\phi)$ parameter space, assuming that the decaying particle accounts for the total dark matter abundance ($r_{\text{DDM}} = 1$). It is straightforward to check that the SNR is linear on $r_{\text{DDM}}$, so the general case can be obtained by rescaling accordingly. The detectable region is defined by the condition $\text{SNR} \geq 8$. Looser conditions (e.g. $\text{SNR} \geq 5$ or $\text{SNR} \geq 1$) lead to slightly larger detectable regions, but the overall shape of the curves remains unchanged. Results are presented for currently operating detectors (LVK, IPTA), instruments expected to become operational within the next few decades (LISA, ET--CE, SKA), and longer-term planned experiments (BBO).
Dotted lines represent the extragalactic contribution, which dominates at higher masses due to redshift effects, while dashed lines correspond to the local contribution, dominant at lower masses. Gray hatched areas indicate existing bounds on the DM lifetime from large-scale structure formation \cite{PoulinEtAl2016} and on the DM mass from dwarf galaxy dynamics \cite{ZimmermannEtAl2025}. Although anisotropy effects on PTA sensitivities have not been included in this analysis, they are expected to enhance the detectability of the signal at masses below the range shown, lying safely beneath the current dwarf galaxy constraints.

\section{Conclusions}\label{sec:conclusions}
In this work, we have studied the detectability of a SGWB produced by the decay of ultralight DM into gravitons. In particular, we have focused on the two graviton decay mode $\phi \rightarrow 2h$ in order to illustrate such phenomenology, although similar qualitative results can be found for a more general decay. Using a model-independent approach, we have provided the first forecasts for the direct detection of this process with current and future GW detectors. As shown in Fig.~\ref{fig:forecast}, this method probes a wide range of lifetimes, many orders of magnitude above the age of the universe and extends down to masses under the lower bound inferred from dwarf galaxy kinematics \cite{ZimmermannEtAl2025}.

The accessible mass range is mainly determined by the frequency band of the experiments. For the detectors considered in this study, $f \sim 10^{-9} - 10^4 \units{Hz}$, corresponding to $m_{\phi} \sim 2 \pi \hbar f \sim 10^{-23} - 10^{-10} \units{eV}$. This range is slightly extended at the high-mass end due to the redshift of the extragalactic component. Although one might expect higher-frequency detectors to probe heavier candidates, Eqs.~\eqref{eq:S_h_E} and~\eqref{eq:SNR_cross} show that the signal-to-noise ratio scales as $f^{-3/2}$, in analogy with single-detector SGWB searches~\cite{Maggiore2007}. Consequently, the detectability of the signal decreases rapidly with frequency. This explains the behavior observed in Fig. \ref{fig:forecast}, where the sensitivity of LISA is over that of CE+ET, despite the latter being much more sensitive in terms of strain. Moreover, as shown in~\eqref{eq:ORF}, when the detector separation exceeds the GW wavelength, the overlap reduction function oscillates rapidly, suppressing correlations and making this method less effective for high-frequency experiments.

In Table \ref{tab:results}, we summarize the detectable range of masses ---considering only masses over the LSS bound, $\tau_{\phi} > 10.2 \,t_0$--- and maximum lifetime for each of the detectors considered in this work, assuming a detection threshold of $\text{SNR} = 8$. The prospects for detecting these decays are expected to improve significantly with the next generation of GW detectors. 
\begin{table}[h]
    \centering
    \begin{tabular}{l l  l}
        \hline \hline
        Detector \hspace{10pt}   & \multicolumn{1}{c}{$m_{\phi}$ [eV]} & \multicolumn{1}{c}{$\tau_{\phi, \text{max}} \, [t_0]$}  \\ \hline
        LVK     & $[1.3 \times 10^{-14}, 1.6 \times 10^{-10}]$ & $5.1 \times 10^{9}$  \\
        ET+CE   & $[1.4 \times 10^{-15}, 3.0 \times 10^{-9}]$ & $2.1 \times 10^{13}$ \\
        LISA    & $[2.9 \times 10^{-20}, 1.1 \times 10^{-12}]$ & $2.1 \times 10^{14}$ \\ 
        BBO     & $[1.3 \times 10^{-18}, 1.4 \times 10^{-9}]$ & $9.9 \times 10^{17}$ \\ 
        IPTA    & $[1.8 \times 10^{-23}, 7.8 \times 10^{-20}]$ & $5.7 \times 10^{9}$ \\ 
        SKA     & $[1.3 \times 10^{-23}, 1.0 \times 10^{-18}]$ & $2.4 \times 10^{13}$ \\ 
        \hline \hline
    \end{tabular}
    \caption{Mass ranges and maximum lifetimes accessible to each detector, assuming a detection threshold of $\text{SNR} = 8$.}
    \label{tab:results}
\end{table}

Over the coming decades, the introduction of LISA, ET, and CE will extend the accessible mass and lifetime ranges, giving a continuous mass coverage from $m_{\phi} \sim 10^{-20} \units{eV}$ to $m_{\phi} \sim 10^{-9} \units{eV}$, and reaching lifetimes up to $\tau_{\phi} \sim 10^{14} t_0$. Further in the future, BBO will provide a wide mass coverage from $m_{\phi} \sim 10^{-18} \units{eV}$ to $m_{\phi} \sim 10^{-9} \units{eV}$, not improving ET+CE bounds on mass, but probing lifetimes up to $\tau_{\phi} \sim 10^{18} t_0$.

Finally, the forecast derived here can be complemented by indirect probes, such as the detection of photons produced through the inverse Gertsenshtein effect  ---the conversion of gravitons into photons in the presence of magnetic fields---, which have been shown to be sensitive to lower lifetimes (close to the LSS bound) but for masses in the $1 - 10^8 \units{GeV}$ scale~\cite{DunskyEtAl2025a}. Future work could also explore the impact of including additional decay channels, such as decays into photons and other Standard Model particles. Although these non-gravitational decay modes would likely dominate the total decay width, including them could open the possibility of studying cross-correlations between gravitational-wave and electromagnetic backgrounds, providing complementary observational handles on decaying dark matter.
\begin{acknowledgments}
 We would like to thank Irene Mariblanca-Escalona for her MSc thesis, which provided the initial motivation for this work. This work is partially supported by the project PID2022-139841NB-I00 funded by MICIU/AEI/10.13039/501100011033 and by ERDF/EU. This work is also part of the COST (European Cooperation in Science and Technology) Actions CA21106, CA21136, CA22113 and CA23130. Additionally, Á.C. acknowledges financial support from MIU(Ministerio de Universidades, Spain) fellowship FPU22/03222.
\end{acknowledgments}

\bibliographystyle{JCAP} 
\bibliography{bibliography}

\end{document}